\documentstyle[epsf]{article}
\begin{document}
\renewcommand{\thefootnote}{\fnsymbol{footnote}}
\begin{flushright}
KEK preprint 94-180\\
NWU-HEP 95-01\\
DPNU-95-02\\
\end{flushright}
\vskip -3cm
\epsfysize3cm
\epsfbox{kekm.epsf}
\begin{center}
{\large \bf
Electron identification using the TOPAZ detector\\
at TRISTAN
\footnote{submitted for publication.}
}\\
\vskip 1cm
Masako Iwasaki$^a$\footnote{internet address: masako@kekvax.kek.jp},
Eiichi Nakano$^b$\footnote{internet address: nakanoe@kekvax.kek.jp},
and
Ryoji Enomoto$^c$\footnote{internet address: enomoto@kekvax.kek.jp}\\
\vskip 0.5cm
{\it
$^a$Department of Physics, Nara Women's University, Nara 630, Japan\\
$^b$Department of Physics, Nagoya University, Nagoya 164, Japan\\
$^c$National Laboratory for High Energy Physics, KEK, Ibaraki 305, Japan\\
}
\vskip 1cm
\end{center}
\begin{abstract}
We present an electron-identification method using the time-projection
chamber and the lead-glass calorimeter in the TOPAZ detector system.
Using this method
we have achieved good electron identification against hadron backgrounds
over a wide momentum range in the hadronic events produced by both
single-photon exchange and two-photon processes.
Pion-rejection factors and electron efficiencies were
163 and 68.4\% for high-$P_T$ electrons
and 137 and 42.7\% for low-$P_T$ electrons in
the single-photon-exchange process, and 8600 and 36.0\%
for the two-photon process, respectively.
\end{abstract}

\section{Introduction}
The TOPAZ detector is located at the TRISTAN $e^+e^-$ collider of KEK.
The center-of-mass energy was 58 GeV and the integrated luminosity
will be more than 300pb$^{-1}$ by the end of 1994.

Electron identification at high purity and high acceptance is
necessary to tag heavy-quark events.
In the $e^+e^-$ annihilation process,
the forward-backward asymmetry of quark-pair
production becomes maximum at the TRISTAN energy region.
A high-accuracy measurement of this was achieved \cite{nagai,nakano}.
In two-photon processes, open-charm production was also studied
with high accuracy \cite{iwasaki}.
Especially for low-$P_T$ charm production, electron
identification provides the best prove for resolved-photon
processes \cite{resolve}, as well as for higher order corrections
of the QCD \cite{nlo}.

In order to carry out these studies we need good electron identification
over a wide momentum range. In particular, the detection of low-momentum
electrons is necessary.
To meet these requirements we have developed a method which uses
an energy-loss measurement with a time-projection chamber and
an energy measurement with a lead-glass calorimeter.
In this article we present related details.

\section{The TOPAZ detector system}
\begin{figure}
\epsfysize18cm
\hskip-0.5in\epsfbox{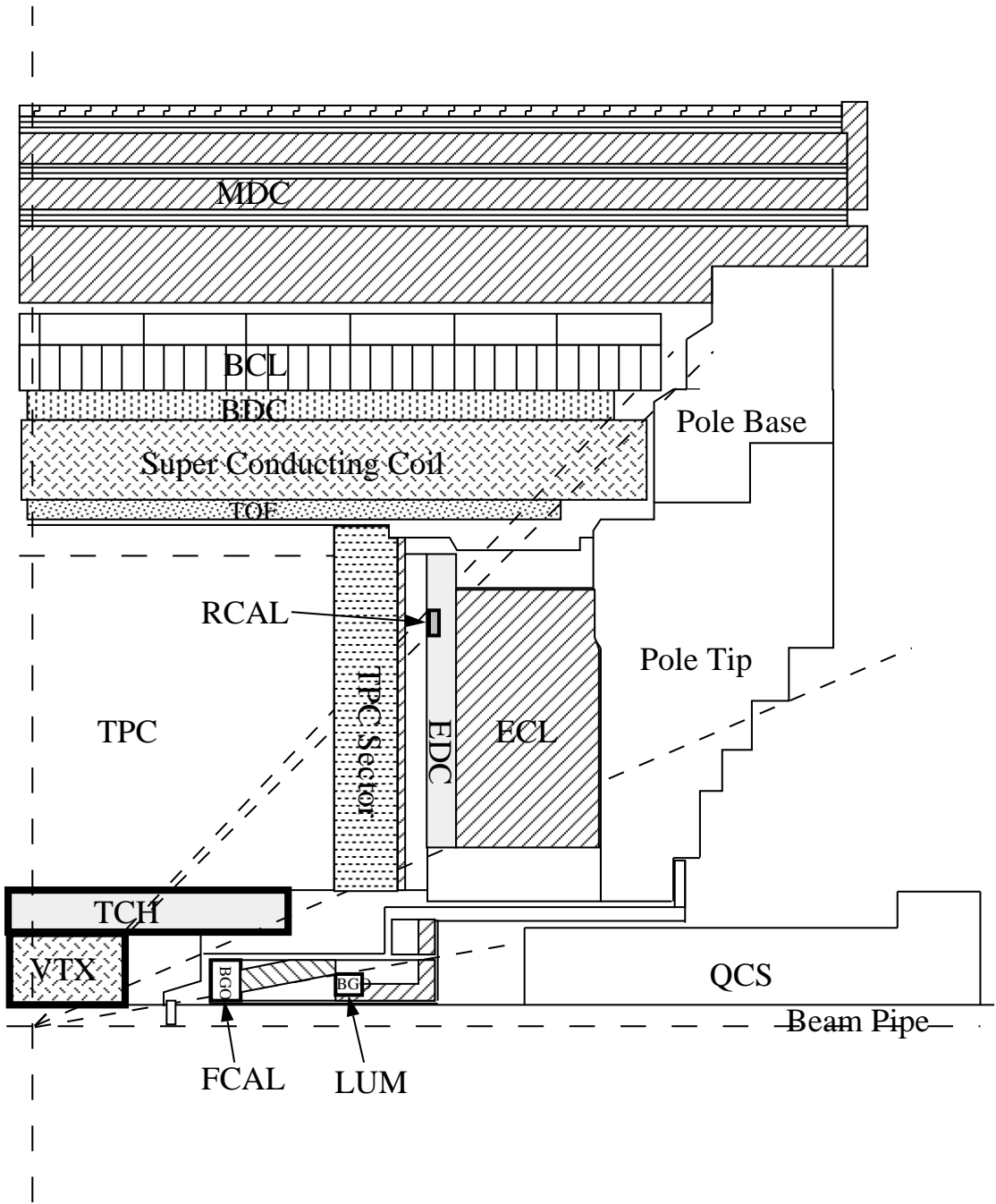}
\caption{
Schematic view of the TOPAZ detector.
}
\label{topazall}
\end{figure}

The TOPAZ detector is a general-purpose 4$\pi$ spectrometer
featuring a time-projection chamber (TPC) as its central tracking
device \cite{tpc,bcl,topaz,trigger}.
A schematic view is shown in Figure \ref{topazall}.
Charged tracks were detected with the TPC
placed in an axial magnetic field of 1.0 tesla.
The energies of the photons and electrons were measured using a barrel
lead-glass calorimeter (BCL)
and an end-cap Pb-PWC sandwich calorimeter (ECL).
A summary of the TOPAZ detector is given in Table \ref{topaztab}.
\begin{table}
\begin{tabular}{lll}
\hline
\hline
Components & Material and size & Measured Performance \\
\hline
Magnet & 2.9m$\phi$ $\times$ 5m, 0.7 $X_0$ & B=1.0 T \\
Beam pipe & 10cm$\phi$, 1mmt Be & \\
Vertex Chamber & Jet-chamber & $\sigma$=40$\mu$m \\
TCH & Cylindrical Drift Chamber & $\sigma$=200$\mu$m \\
TPC & 2.4m$\phi$ $\times$ 2.2m & $\sigma_{r\phi}$=185$\mu$m \\
    & 16 sectors & $\sigma_z$=335$\mu$m \\
    & 10 pad rows, 176 wires / sector & $\sigma_{dE/dx}$=4.6\% \\
    & Ar/CH$_4$ (90/10) 3.5atm
       & $\sigma_{P_T}/P_T=\sqrt{(1.5P_T)^2+1.6^2}\%$ \\
TOF & Plastic Scintillator & $\sigma$=220ps \\
BDC & Streamer tube & $\sigma$=350$\mu$m \\
BCL & Lead-glass (SF6W), 20$X_0$
       & $\sigma_E/E=\sqrt{(8/\sqrt{E})^2+2.5^2}\%$ \\
    & 4300 blocks & $\sigma_{\theta}$=0.38$^o$ \\
    & $|\cos \theta|\leq$0.82 & \\
ECL & Pb-PWC sandwich & $\sigma_E/E=6.7\%$ \\
    & 18 $X_0$ & (for 26-GeV electron) \\
    & 0.85$\geq |\cos \theta|\leq$0.98 & $\sigma_{\theta}$=0.7$^o$ \\
\hline
\hline
\end{tabular}
\caption{Performance of the TOPAZ detector components.}
\label{topaztab}
\end{table}
The coordinate system used in this article is as follows: z is defined
as the beam-electron direction; x is the vertical axis
and y is the horizontal axis.
The major components of the TOPAZ detector are described in
the following sections.

\subsection{The Time-Projection Chamber (TPC)}
The Time-Projection Chamber (TPC) is a three-dimensional tracking
chamber which can identify particle species by energy-loss
measurements. It has a fiducial volume of 2.4m in diameter
and 2.2m in axial length, comprises 16 multiwire proportional
counters (sectors), each equipped with 175 sense wires and 10 rows
of segmented cathode pads, and is filled with a gas mixture of 90\%
Ar and 10\% CH$_4$ at 3.5 atm \cite{tpc}.

Charged particles passing through the TPC sensitive region ionize
gas molecules and liberate electrons along their trajectories.
These electrons drift along the direction of the electric field toward
the sectors at a speed of 5.3 cm/$\mu$sec,
and finally produce avalanches around sense wires; induced signals
are detected by cathode pads.
In order to reduce any space-charge effects, the gating-grid plane is placed
above the shielding-grid plane.
The signals from wires and pads are amplified and are shaped by
analogue electronics. They are then fed into charge coupled
device (CCD) digitizers sampling the wave-forms in units of a 100nsec
``bucket" from which the z-positions and energy-loss ($dE/dx$) information
are extracted.

The spatial and momentum resolutions are studied with cosmic rays
as well as
Bhabha and $\mu^+\mu^-$ events.
The spatial resolution was obtained to be $\sigma_{xy}$=185$\mu$m
and $\sigma_z$=335$\mu$m.
The momentum resolution was obtained by comparing two measurements in the
opposing sectors in cosmic-ray events to be
$$\sigma_{P_T}/P_T=\sqrt{(1.5P_T)^2+1.6^2}\%.$$
The $dE/dx$ resolution was studied using minimum-ionizing pions in the
beam events. The obtained resolution is $\sigma_{dE/dx}=$4.6\% after
calibration (discussed later).

\subsection{The Barrel Lead-Glass Calorimeter (BCL)}
The Barrel Calorimeter (BCL), which comprises 4300 lead-glass \v Cerenkov
counters, has a cylindrical structure covering an angular region of
$|\cos\theta|\leq0.82$. The calorimeter is divided into 72 modules,
8 in the $\phi$ direction and 9 in the z direction, each of which  has
an array of 60 lead-glass counters \cite{bcl}.

Each lead-glass counter is made of SF6W and has 20 radiation
lengths, so that it can absorb more than 95\% of
the shower energies at TRISTAN's highest energy. The lead-glass
blocks are tilted by 1.8$^o$ with respect to the radial line
in the $r-\phi$ plane so that no photons coming from the
interaction point escape through the counter-to-counter gaps.
The \v Cerenkov light emitted in a lead-glass counter is detected by a
photomultiplier after passing through a light guide. The signal from
the photomultiplier is sent to a digitizer and is used for a neutral
trigger \cite{trigger}. The system is calibrated by using
a Xe flash lamp.

The performance of the BCL has been studied by using Bhabha events.
The energy and spatial resolution were determined to be
$\sigma_E/E=\sqrt{(8/\sqrt{E})^2+2.5^2}$\% and $\sigma_{\theta}$=0.38$^o$,
respectively.

\section{Track reconstruction}

The raw data was first corrected
for any channel-to-channel variation of electronics. The spatial
positions of ``hits" in the TPC were then determined using pad signals, and
particle tracks were searched among theses points. Each found
track was fitted to a helix, and the charge and momentum were assigned.

\begin{figure}
\epsfysize8.4cm
\epsfbox{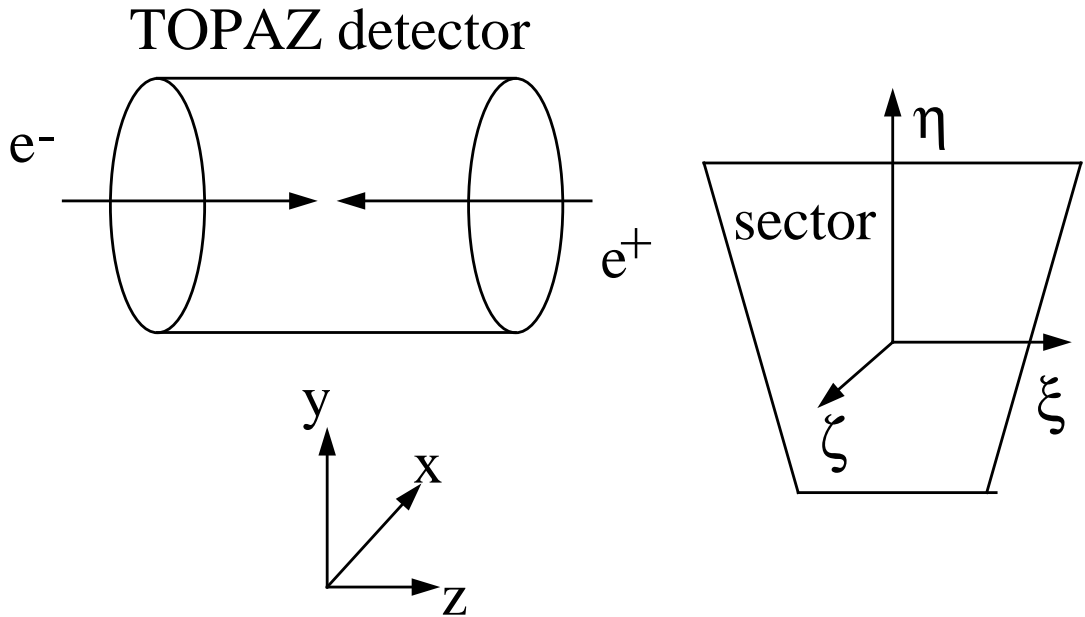}
\caption{
Definition of the coordinate system in the TOPAZ detector.
}
\label{coodtpc}
\end{figure}

	We define the coordinate system in the TOPAZ detector
as shown in Figure~\ref{coodtpc}.
The global coordinate in the TOPAZ detector is defined by
$x$, $y$ and $z$. The origin of the coordinates is the detector center.
The $\xi$, $\eta$ and $\zeta$ are coordinates fixed on the TPC sector.
The $\xi$-axis is
along the sense-wire direction, the $\eta$-axis points in the
radial direction at the center line of the sector, and the $\zeta$-axis
is along the beam line.
The details concerning this section can be found in reference \cite{kfujii}.

\subsection{Space-point reconstruction}

	The pad signal of each channel was first corrected for any electronics
variation, and was then examined as a function of the CCD bucket number.
Contiguous signals were grouped into a cluster to form a $\zeta$ hit. The
peak pulse height and the $\zeta$ position of the cluster were calculated
by fitting the highest three CCD buckets to a parabola. Clusters were
then searched in the $\xi$ direction. If signals in contiguous pads had
roughly the same $\zeta$ positions (within 1cm), theses signals were
combined into a cluster.

	When a cluster was found, the $\xi$ and $\eta$ positions of the
cluster were determined. Since the charge distribution along the pad row
has approximately a Gaussian shape, the $\xi$ position was calculated by
fitting the highest 3 pad pulse heights to a Gaussian. If the cluster had
only two pad signals, the pre-determined width of the Gaussian was used in
the fit. The $\eta$ position of the cluster was calculated by taking the
centroid of the signals of five wires nearest to the pad row.

	The position
resolutions in $\xi$ and $\zeta$ were studied using cosmic-ray tracks;
typical values were obtained as $\sigma_{\xi} = 230 \mu m$ and
$\sigma_{\zeta} = 340 \mu m$.

\subsection{Track reconstruction}

To find tracks from the reconstructed hits in the TPC,
all found space points were first filled in a 2-dimensional
histogram with $\phi$ and $z(R_{ref})$ axes such that
\begin{eqnarray*}
  \phi & = & {tan}^{-1}(x/\sqrt{x^2+y^2}) \\
   z(R_{ref}) & = & z \times ( R_{ref}/R ),
\end{eqnarray*}
where $R_{ref}$ is the radial position of the fixed reference point.
The space points associated to a track from the origin with a relatively
high momentum were clusterized in a bin of the histogram.
The track finding began by searching for a bin containing more space points
than a given threshold.
The threshold was set at 9 in the beginning,
which corresponded to 10 pad raw hits;
it was then lowered by one when all of the bins
above the current threshold were
found. Lowering of the threshold was repeated
until the threshold became two (three pad rows).

	To test whether the space points in the bin formed a track or not,
three space points which spanned the
largest lever arm were chosen and checked
as to whether they were on a single helix.
If the space points were on a helix, other space points in the bin
were checked as to whether they were consistent with the helix. If not,
another set of three space points were chosen and the test was
repeated. If a sufficient
number of space points could be associated they were
recognized to form a track and were fitted to a helix function.

	Before the helix fitting, the space points were corrected for
any dependence on the track angle to the sectors.
The helix function was parametrized as follows:
$$x = \frac{1}{\kappa}(cos\phi_0-cos(\phi+\phi_0)) + d_r cos\phi_0 + X_0$$
$$y = \frac{1}{\kappa}(sin\phi_0-sin(\phi+\phi_0)) + d_r sin\phi_0 + Y_0$$
and
$$z = -(\frac{1}{\kappa})\phi tan\lambda + d_z + Z_0,$$
where
\begin{description}
\item $(X_0, Y_0, Z_0)$ is the position of the pivot,
\item $\kappa$ is the inverse of the radius of the helix,
\item $\phi_0$ is the angle of the pivot to the $x$ axis,
\item $d_r$ and $d_z$ are the distance of
       the pivot from the true trajectory, and
\item $\lambda$ is the the angle which the track makes with the $xy$ plane.
\end{description}
The parameters to be determined by the fit are
\begin{center}
\[ d_r, \phi_0, \kappa, d_z, {~and~tan}\lambda. \]
\end{center}
The fitting of the helix was achieved by minimizing $\chi^2$
defined as
$$\chi^2 = \sum_{i} ( (  \frac{\xi_i -
                    \xi(\eta_i)}{\sigma_{\eta_i}} )^2 +
                    ( \frac{z_i - z(\eta_i)}{\sigma_{z_i}} )^2 ),$$
where $\xi_i$ and $z_i$ are the measured position in the i'th pad row and
$\eta(\eta_i)$ and $z(\eta_i)$ are the expected position by the helix.
We checked this $\chi^2$ and made
a final decision as to whether to recognize the space points
as a track or not.

	After all histogram bins forming tracks were found, the
space points which were not recognized as tracks were then examined.
{}From the remaining histogram bins, the bin containing the largest
number of space points was searched for. The inner-most space point in the bin
was chosen to be the origin (pivot) of the new histogram.
The new histogram was defined with wider bins so that the space points of
a low-momentum track
could be clusterized in a bin as a new track candidate.
The bins were tested in the same way as described above.

\subsection{Track Refinement}

	Although the gating grid prevents the feed-back of positive ions into
the drift space and reduces the distortion in the electric field in the TPC,
a small amount of distortion still remains,
resulting in a systematic shift in
the space-point determination. The dielectric material in the TPC,
such as field cages, are also a source of distortion.

	In addition to electrostatic distortion, the spatial positions of
the TPC sectors may deviate from the designed position. This also causes a
shift in the determination of the space points.

	The electrostatic distortions in the
$r\phi$ plane were parametrized as
$$\Delta x_\phi = a_0 ( r - r_{ref} )^2,$$
where $\Delta x_{\phi}$ is the position displacement
in $\phi$ due to the distortion.
The deviations of the sector position in the $r\phi$ plane
($\Delta\zeta$, $\Delta\eta$)
are parametrized as
$$\Delta\xi = d_\xi - \delta\eta$$
and
$$\Delta\eta = d_\eta + \delta\xi.$$
The shift in the $z$ direction ($dz$) was expressed
so as to include both the electrostatic
distortion and the position displacements of the sectors, and
was parametrized as
$$\Delta z = d_z + d_vz.$$
These parameters were defined for each sector.
The parameters of the electrostatic distortions were determined
based on a
comparison of two cosmic-ray data taken when the beam was on and off.
The parameters of the position displacements of the
sectors and the shift in the
$z$-direction were determined based on a
comparison of two measurements for a cosmic-ray track by two sectors
when the beam was on.
The space points calculated in the previous sections were
corrected for any distortions using these parameters.

	Finally, the space points were fitted to a helix again
in order to assign the
sign of the charge and momentum to the track.
The momentum in the $xy$ plane ($P_T$) and
along the $z$-axis ($P_L$) were expressed using the helix parameters as
$$P_T = |\frac{Q}{\alpha\kappa}|$$
and
$$P_L = P_T tan\lambda,$$
where
$$\alpha = \frac{1}{cB} = 333.56\;cmGeV^{-1}.$$
$Q$ is the charge of the particle, $c$ the velocity of light,
and $B$ the strength of the magnetic field (1.0 tesla).
The sign of the charge was determined from the sign of $\kappa$.

	The resolution of the momentum measurement
was estimated using the cosmic rays \cite{shirahashi}. The resolution was
studied by comparing two measurements in opposite sectors for a
cosmic-ray track.
The momentum resolution was studied as a function of
momentum,
and was obtained as
$$\frac{\Delta P_T}{P_T} = \sqrt{[(1.5\pm0.1)P_T]^2 + (1.6\pm0.3)^2} \%.$$

	The effect of distortion corrections was studied using Bhabha
events.
The momentum resolution for the Bhabha events was measured to be
$(1.7 \pm 0.1)P_T$\%.

\section{Energy-loss measurement in the TPC}
The energy loss per unit length ($dE/dx$) is a function
of the velocity ($\beta$) and the charge ($Q$) of a particle.
In this section we examine the property of the $dE/dx$ curve
and discuss what determines the $dE/dx$ resolution.
In practice, the $dE/dx$ resolution suffers from various
systematic shifts due to temperature, pressure, high voltage, electronics
and so on.
We need to improve the resolution as much as possible
in order to maximize
the searchable region.
A detailed description of the derivation of $dE/dx$ is given here.
\subsection{Theoretical consideration}

\subsubsection{Average energy loss}
       When a charged particle traverses a medium it interacts with atoms
in the medium.
The dominant process of energy loss is atomic excitation.
The probability of exciting an atom scales as the square of the transverse
component of the incident particle's electric field with respect to its
velocity vector.
The transition probability, and hence the energy loss,
scales as the charge of the incident particle squared.

     Qualitatively, the average energy loss ($<dE/dx>$)
in the material can be described
as a function of the velocity ($\beta$)
and charge ($Q$) of the particle.
The function can be
approximately written as
 $$ <\frac{dE}{dx}> \approx \frac{4\pi nQ^2e^2}{m_ee^2\beta^2}
                 ( ln \frac{2m_e\beta^2\gamma^2}{I^2 +
                            (\hbar\omega)^2\beta^2\gamma^2} - 2\beta^2),$$
where
\begin{description}
  \item $I$ is the logarithmic mean atomic ionization potential,
  \item $\omega$ is the plasma frequency of the medium,
  \item $n$ is the electron density of the medium,
  \item $m_e$ is the electron mass,
  \item $e$ is the electron charge, and
  \item $Q$ is the charge of the incident particle.
\end{description}
The values of $<dE/dx>$ fall as
1/$\beta^2$ with increasing the momentum of the particles.
The fall off is due to the fact
that a particle traversing the material spends
less time in the electric field of atoms, and, hence, transfers less
energy as the particle's velocity increases.
The ionization becomes minimum at $\beta\gamma \simeq 3$,
and then $<dE/dx>$ rises
logarithmically as the momentum increases. This ``relativistic rise"
is the result of the relativistic increase in the transverse electric
field of the incident particle. The $<dE/dx>$ finally reached the
plateau due to the ``density effect" caused by the polarization of
the medium.

\subsubsection{Energy-loss distribution}
      There are two contributions to the shape of the energy-loss
distribution: atomic excitations and  $\delta - ray$ production.
The $\delta-ray$ contribution falls inversely
proportional to the energy-transfer
squared (1/$(\Delta E)^2$), because this is the classical Rutherford
scattering for free electrons.
This process makes a long tail, called Landau tail, in the energy-loss
distribution up to the kinematical limit.
The peak of energy loss distribution is due to
atomic excitation, which is the dominant energy-loss process.
Argon, which is the main component of the TPC gas,
has three shell structures: K, L, and M.
The energy levels are 3.20 keV, 248 eV, and 16 - 52 eV, respectively.
Since the collision probability with an electron is roughly
proportional to its energy level, the lower energy-level
excitations give a narrower $dE/dx$ distribution.

\subsection{$dE/dx$ measurement}
\subsubsection{Derivation of the energy loss}
	The energy loss was measured using wire signals of the TPC.
First, the wire signal clusters were searched in the $z$ direction
in the same way as the pad clusters.
The pulse height of a cluster was determined from the peak
amplitude of a parabola fitted to the highest 3 CCD buckets in the cluster.
The $z$ position of the cluster was also determined from the fit.
The cluster was then associated to a track using the $z$ information.
If the cluster was within 1 cm of the track in the $z$ direction, and not
associated to other tracks,
the cluster was linked.

	The pulse-height amplitude was converted into the energy-loss
value by two steps.
The first was a correction for any non-linearity of the electronics.
The second was a correction for the wire-to-wire gain variation.
Details concerning the calibration are described in the following subsection.

	Since a TPC sector has 175 sense wires,
the energy loss for a track was sampled every 4-mm segment
up to 175 independent data.
A single variable, which is similar to the most probable energy loss, was
deduced from the data sample.
The final $dE/dx$ was then derived by making various corrections, which
are discussed later.

\subsubsection{Calibrations}
	Prior to the experiment, after the electronics
were calibrated by pulsing
the shielding grid wires with various test-pulse amplitudes,
the calibration curve for each channel was determined.
The linearity of our electronics is  better than 0.5\%
below the saturation point.
The remaining non-linearity, appreciable near to the saturation point,
is corrected using this calibration curve.

	After an electronics calibration, a correction for
the wire-to-wire gain variation is made.
Each sector is equipped with three rods having a $^{55}Fe$ X-ray
source for each wire. Each rod can be moved pneumatically behind the
hole on the cathode plane to irradiate wires. The wire-to-wire gain
variation is corrected by calibrating the pulse height for the main peak
of 5.9 keV for each wire. This calibration reduces the gain variation over
a sector within a 3\% level, and also gives the factor
necessary to obtain the absolute
value of the measured $dE/dx$.

\subsubsection{The 65\%-truncated mean}
\begin{figure}
\vskip -2cm
\epsfysize10cm
\hskip1in\epsfbox{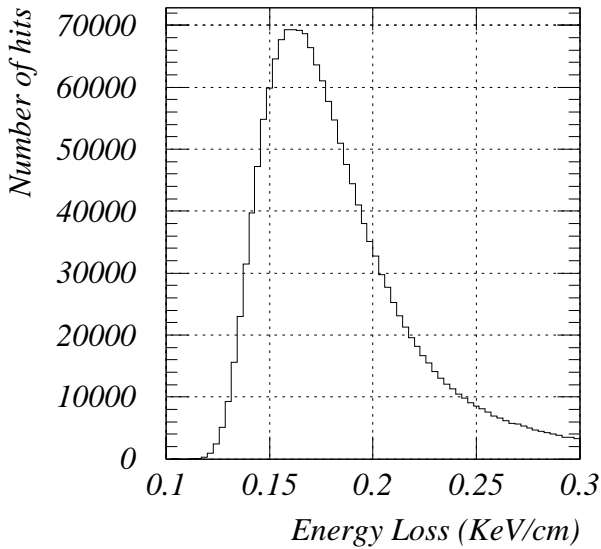}
\vskip -1cm
\caption{
Pulse-height distribution obtained by the TPC wire signals
for minimum ionizing pions.
}
\label{Landau}
\end{figure}

	Figure~\ref{Landau} shows the
measured energy-loss distribution in the
TPC for minimum ionizing pions.
The shape is asymmetric and
has the long tail at the higher side.
Therefore, a simple mean of the $dE/dx$ sample for a track is not
a good parameter for particle identification.
The most probable energy loss or the mean value around the peak
is a better parameter. We adapted the mean of the lowest 65\%
in the data sample (the 65\%-truncated mean) as an estimator for
the energy loss of a track.
In the following we use $dE/dx$ as in this meaning.

\subsubsection{Corrections}
      To achieve a good resolution in $dE/dx$ measurements,
one must reduce the systematic shifts of the gain.
In this section we discuss the systematic shifts of $dE/dx$
and how we correct them.

\begin{figure}
\vskip -2cm
\epsfysize10cm
\hskip1in\epsfbox{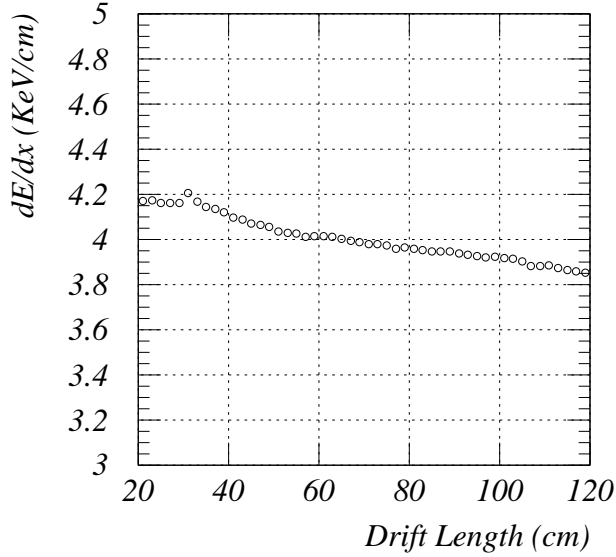}
\vskip -1cm
\caption{
$dE/dx$ as a function of the drift length.
}
\label{DEDX attenuation}
\end{figure}

      In the TPC, measuring the energy loss is an indirect process.
An implicit assumption is that the number of ionization electrons
produced per track length is proportional
to $dE/dx$ of the charged particle.
As electrons in the gas drift to the sectors
they defuse; some fraction is absorbed
by impurities, such as $O_2$, in the gas.
Since the maximum drift length is more than 1 m, this effect
is not negligible.
We determined the attenuation factor using minimum ionizing pions
in the momentum range between 0.5 and  0.6 GeV/c.
Figure~\ref{DEDX attenuation} shows the measured value of
$dE/dx $ as a function of the drift length.
We correct this effect using following function:
 $$   PH_{correct} = PH_{measured} \times( 1 + C_1L ),$$
where
\begin{description}
  \item $PH$ is the pulse height, and
  \item $L$ is the drift length.
\end{description}
The attenuation factor ($C_1$) is typically 7\% per meter.

\begin{figure}
\vskip -2cm
\epsfysize10cm
\hskip1in\epsfbox{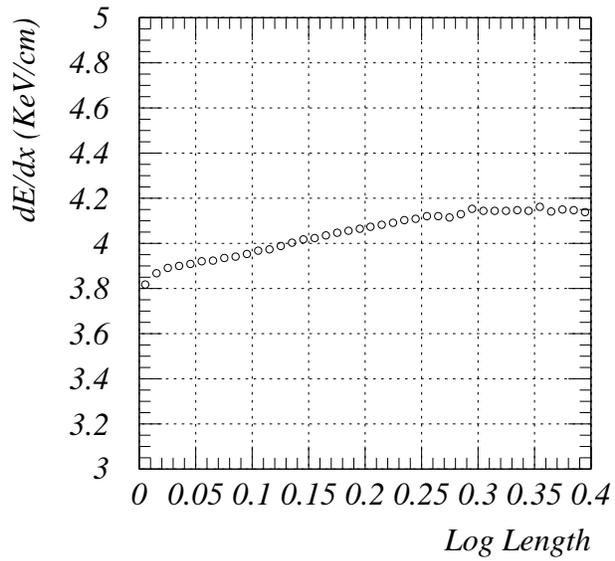}
\vskip -1cm
\caption{
$dE/dx$ as a function of the sample thickness in
logarithm after a segment-length correction.
}
\label{DEDX log}
\end{figure}

	The gain shift due to the sample thickness dependence
was also corrected using the minimum ionizing pions.
Figure~\ref{DEDX log} shows $dE/dx$ vs. the path length on the logarithm
scale.
This systematics was corrected using
 $$   PH_{correct} = PH_{measured} \times ( 1 - C_2\times log(X) ),$$
where
\begin{description}
  \item $X$ is the path length normalized by 4 mm.
\end{description}
The factor $C_2$ was determined to be 0.2.

      The electric fields near to the sense wires also affect
the avalanche process.
The dependence of the pulse height on the high voltage was
parametrized as
 $$   PH_{correct} = PH_{measured} \times ( 1 - C_3\times
                   \frac{\delta V}{V_0} ),$$
where
\begin{description}
  \item $V_0$ is the nominal high voltage (1970V) on the sense wire, and
  \item $\delta V$ is the variation of the voltage.
\end{description}
The correction factor ($C_3$) of the pulse height was determined
using data taken in low high-voltage cosmic-ray runs.
A 1\% high-voltage change corresponds to a 17\% change in the gain.

\begin{figure}
\vskip -2cm
\epsfysize10cm
\hskip1in\epsfbox{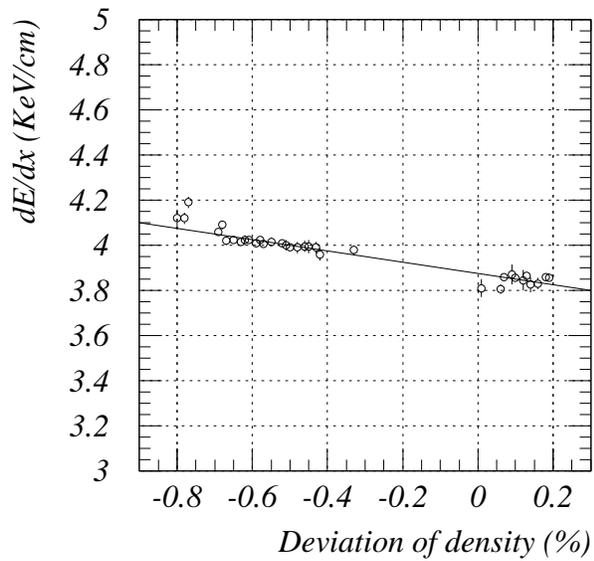}
\vskip -1cm
\caption{
$dE/dx$ as a function of the density from the nominal value.
}
\label{DEDX temperature}
\end{figure}

      The electrons approaching the sense wires
pass through the avalanche process.
As the gas density increases
the mean-free path of electrons becomes shorter and
the soft collisions in which the electrons do not ionize the atoms increase;
hence, the amplification factor drops.
The gas density is proportional to the pressure and inversely proportional to
the temperature.
The pressure was well controlled, and its variation was less than 0.04\%
throughout this experiment.
The temperature of the gas was increased by about $2^\circ$C during
the summer.
Figure~\ref{DEDX temperature} shows $dE/dx$ vs. the variation in the
gas density.
{}From the temperature change, we found that
 $$   PH_{correct} = PH_{measured} \times
                    ( 1 + C_4\times\frac{\delta \rho}{\rho_0} ),$$
where
\begin{description}
  \item $\rho_0$ is the nominal gas density, and
  \item $\delta \rho$ is the variation of the gas density.
\end{description}
The $C_4$ value was 6.0.
A $1^\circ$C temperature
change corresponds to about a 2\% change in the gain.

After these corrections, we again carried out a wire-to-wire gain calibration.
The event sample with minimum ionization pions were used for this
purpose. After the pulse-height distributions for each wire were made and
the correction factors were derived, a run-dependent correction
was carried out. There were still unknown factors which changed the
energy-loss values. Each step used a single run corresponding to
a few 1000 beam events.

Finally, we derived the $dE/dx$ distribution
for momentum between 0.5 and 0.6 GeV,
which was the minimum ionizing pion region. There were two peaks
corresponding to pions and electrons.
For several reasons, the relationship
between these two peak values changed nonlinearly.
One reason was an over-value setting of the threshold values in the
electronics during some experiments.
The other was a selection bias in the event pre-selection code
in order to reduce the data size, mainly due to the vacuum condition.
These caused a nonlinear offset in the pion
$dE/dx$ value. We thus corrected these by normalizing them to the initial
experimental values in which the threshold values were considered to be
sufficiently low.
Furthermore, we fitted the width of two $dE/dx$ peaks using the
common ratio to the center values of the peaks using two Gaussians. The
fitted width was considered to be the effect of this nonlinearity; the
$\chi^2$s calculation is described later.

\subsubsection{Velocity dependence of $dE/dx$}
\label{bethe}
\begin{figure}
\vskip -2cm
\epsfysize10cm
\hskip1in\epsfbox{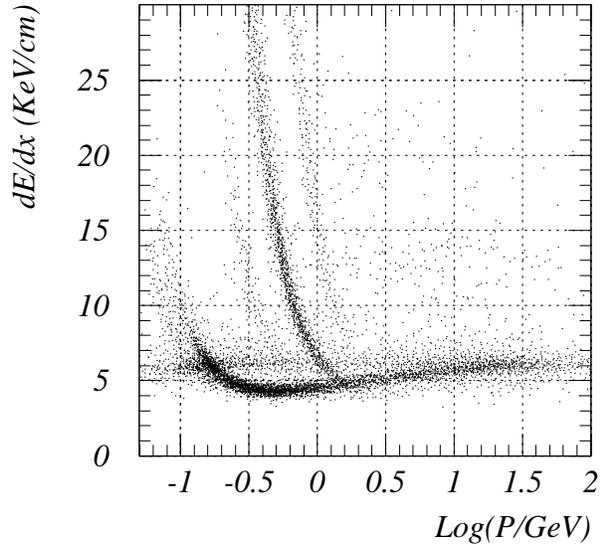}
\vskip -1cm
\caption{
$dE/dx$ vs. momentum for tracks in more-than-one-track events.
}
\label{TPCPLOT}
\end{figure}

	The $dE/dx$ curve as a function of
the particle velocity was determined from
the data. Although theoretical calculations of the $dE/dx$ curve
reproduce the data fairly well, they are not adequate for
particle-identification purposes, especially in the relativistic rise region.
The data used for the determination were those from
protons for the small-$\beta$ region,
cosmic-ray muons and minimum
ionizing pions for low $\beta\gamma$, and electrons for the
relativistic rise region. The data are fitted to a ninth-order polynomial
parametrized as
$$
\frac{dE}{dx} = \frac{1+(\beta\gamma)^2}{(\beta\gamma)^2}
                \sum_{i=0}^{9} c_i ( \frac{ln(\beta\gamma)-4}{5})^i
		-0.70659,
$$
where the $c_i$ are given in Table \ref{coefdedx}.
\begin{table}
\begin{center}
\begin{tabular}{cc}
\hline
\hline
coefficient & value \\
\hline
$c_0$ & 6.2133   \\
$c_1$ & 2.1340   \\
$c_2$ &-1.6116   \\
$c_3$ &-0.19783  \\
$c_4$ & 0.0363802\\
$c_5$ &-0.35785  \\
$c_6$ & 1.4957   \\
$c_7$ &-0.16234  \\
$c_8$ &-0.95077  \\
$c_9$ & 0.379    \\
\hline
\hline
\end{tabular}
\end{center}
\caption{Coeficients of the 9'th-order polynomial function.}
\label{coefdedx}
\end{table}
	Figure~\ref{TPCPLOT}
shows the $dE/dx$ vs. momentum for tracks in
more-than-three-track events.
The bands of electrons, pions, kaons and protons can be clearly seen.

\subsubsection{$dE/dx$ resolution}
\begin{figure}
\vskip -2cm
\epsfysize10cm
\hskip1in\epsfbox{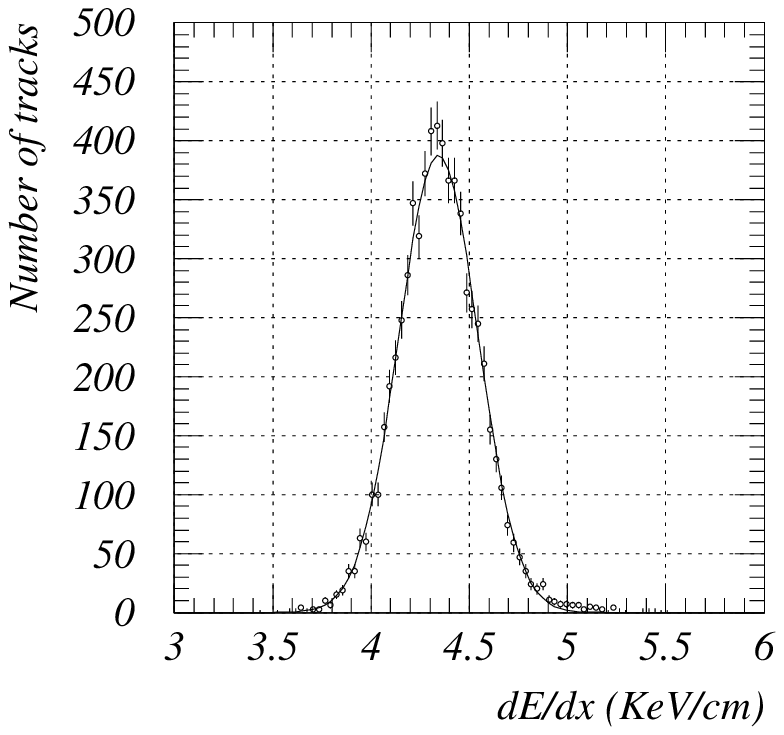}
\vskip -1cm
\caption{
$dE/dx$ distribution for minimum ionizing pions.
}
\label{DEDX resolution}
\end{figure}

        The resolution of $dE/dx$ was studied using the
minimum ionizing pions. Figure~\ref{DEDX resolution} shows
the $dE/dx$ distribution of pions with momenta between 0.5
and 0.6 GeV/c.
The resolution was estimated by fitting a Gaussian function
to the distribution, and was obtained as
$\sigma_{dE/dx} = 4.6 \%$. The resolution for Bhabha events
was also measured as
$\sigma_{dE/dx} = 4.4 \%$.

\section{Energy measurement in the BCL}
\subsection{Clustering}
The electromagnetic clusters detected with the BCL were used for electron
identification. The method for electromagnetic energy clustering
is described as follows:
\begin{enumerate}
\item We searched the maximum energy counter among 8 neighboring
counters (at the +z or -z edge of the BCL, 5 counters) for counter by
counter;
\item If the neighboring maximum counter energy was higher than its
energy, the counter was merged to the neighboring one;
\item After the above procedures were iterated, the clusters were formed;
\item The cluster energy was calculated by summing the total
counter energy in
the cluster;
\item The azimuth angle ($\phi$) and the z-position were calculated using
the energy-weighted mean of the counter positions;
\item The depth of the cluster center was calculated according to
$X=ln(E/E_c)+1.5$ (radiation length),
where $E$ is the cluster energy and $E_c$ (12.6 MeV) is the
critical energy of the BCL (lead glass);
\item The radial position (R) of the cluster was calculated
according to the formula
$R=R_{BCL}+X_0\cdot X\cos\theta$, where
$R_{BCL}$ and $X_0$ are the inner radius and the unit radiation length
of the BCL, i.e. 176cm and 1.7cm, respectively. Here, $\theta$
was assumed to be that of the maximum energy counter in the cluster, at first;
\item From z and this R, the $\theta$ was re-calculated; and
\item R and $\theta$ calculations were iterated two times.
\end{enumerate}

\subsection{Energy calibration}
The monitoring system for the gains of 4300 counters comprises just
one light source and an optical-fiber system.
As the light source, a xenon flash lamp having a 3mm gap was chosen, because
the emission spectrum resembles \v Cerenkov light. The amount
of light received at a counter corresponds to about 7 GeV in energy.
In order to correct for any variation in the light
intensity of the xenon flash lamp, reference counters
were placed in a temperature-controlled box.
At each reference counter a NaI(Tl)-$^{241}$Am light pulser
was attached to measure the gain independently.
The gain variation can be corrected to the 0.5\% level.

The gain of each counter was calibrated with 4, 2, and 1 GeV electron
beams from the IT1 beam line \cite{it1} at the 8-GeV Accumulator
Ring (AR) at KEK.

We have accumulated Bhabha scattering events, after standard cuts
of requiring two clusters of deposited energy greater than
1/3 of the beam energy at $|\cos\theta|<0.77$
satisfying an acolinearity angle being less than 10$^o$.
The obtained resolution was 4.5\%, apparently dominated by the
nonuniformity in the gains of individual counters.
The angular resolution
was obtained to be 0.36$^o$.
Calibration of the individual counters was carried out using these Bhabha
events. However, not all of the counters have been calibrated yet, due to a
lack of statistics.

The electron energy-loss in the material in front of the BCL was
investigated by studying the E/P for electrons.
In this study we used an experimental two-track sample selected by
the following criteria, expecting these to be dominantly
$e^+e^-\rightarrow e^+e^-e^+e^-$ events:
\begin{enumerate}
\item The number of charged tracks with $P_T>0.15$ GeV, the
polar angle $|\cos\theta|<0.83$, and the number of degrees of freedom
(N.D.F.) of charged-track
fitting $\geq$ 3 had to be 2;
\item The number of the BCL clusters with $E_{cluster}>100$ MeV had to be
at least 1; and
\item The visible energy ($E_{vis}$) of the event had to satisfy
$E_{vis}<30$ GeV, where both the charged tracks in the TPC
and the clusters in the BCL were used.
\end{enumerate}
Most of the selected events were expected to be two-photon events
of ($e^{+}e^{-}$)$e^{+}e^{-}$, ($e^{+}e^{-}$)$\mu^{+}\mu^{-}$, and
($e^{+}e^{-}$)$\pi^{+}\pi^{-}$.
The remaining ones were $\mu^{+}\mu^{-}$, $\tau^{+}\tau^{-}$,
$e^{+}e^{-}\gamma$, etc.
The used data corresponded to an integrated luminosity of 24.5 pb$^{-1}$ at
$\sqrt{s} = 50 \sim 60.8 $ GeV.
These selection cuts left 9120 two-track events.

To separate electron tracks from muons and pions, we used
energy-loss ($dE/dx$) information from the TPC and TOF as follows: (1)
The electron-candidate tracks were required to have $dE/dx$ in the range of
5.5 $\leq$ dE/dx $\leq$ 7.5 keV/cm;
(2) the confidence level (CL) for the electron hypothesis by the TOF
information had to be $>$ 0.01.
We then calculated the E/P ratio of the energy measured by the BCL and
the momentum measured by the TPC. The TPC-BCL combination was chosen
when the distance between the BCL cluster and the extrapolated track
was shortest.
We only calculated the E/P for the TPC-BCL combination whose distance
was less than 10cm.

\begin{figure}
\vskip -2cm
\epsfysize10cm
\hskip1in\epsfbox{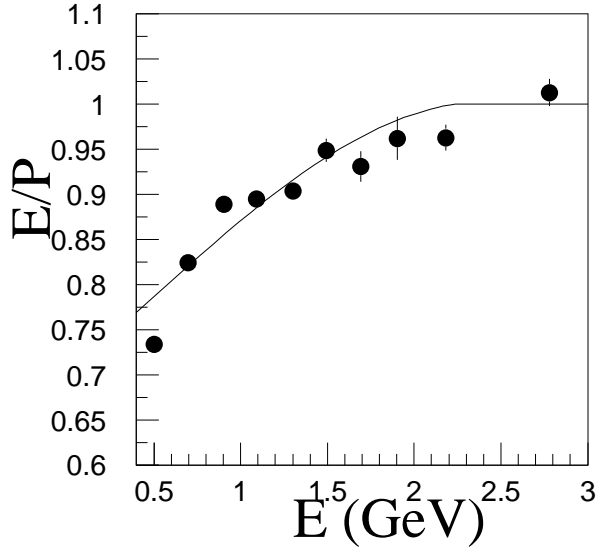}
\vskip -1cm
\caption{
E/P ratio as a function of the BCL energy (filled circle).
The solid line was
obtained from a polynomial fitting (described in the
text).
}
\label{fig:fitting}
\end{figure}

\begin{figure}
\epsfysize18cm
\epsfbox{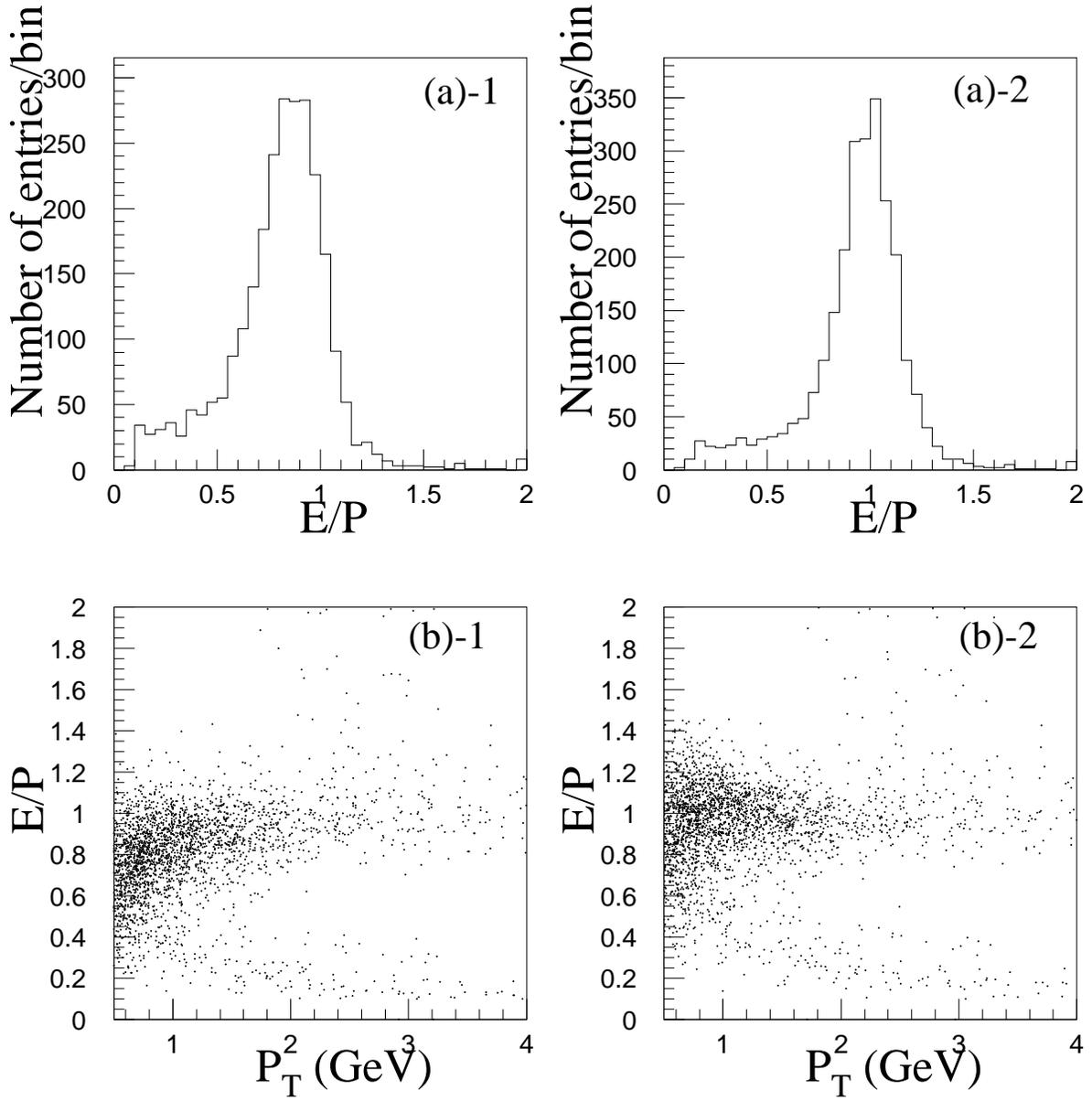}
\caption{
E/P distributions in the momentum range $0.4\leq P\leq 4.0$ GeV
[(a)-1 and 2] and
Scatter plots of E/P vs. energy [(b)-1 and 2].
Indices 1 and 2 denote that without and with the BCL correction,
respectively. The equation of the correction function is given in the
text.
}
\label{fig:E/PvsP}
\end{figure}

To determine the BCL energy correction we fitted
the E/P distribution with a second-order
polynomial of the BCL
energy.
We used electron candidates which were in the energy range E $<$
4.0 GeV
and the polar angle range $|\cos\theta| \leq$0.77.
The E/P-versus-energy distribution with the fitting curve is shown in
Figure \ref{fig:fitting}.
Since real electrons should concentrate on E/P = 1, we obtained the
following expression from the fitting result:
\begin{eqnarray*}
    \frac{\Delta E}{E}&=&0.430 - 0.354\cdot E + 0.072\cdot E^2
\;\;\;\; for \;\; E\leq 2.3 GeV,~and \\
  &=& 0 \;\;\;\; for \;\;E > 2.3 GeV.
\end{eqnarray*}
The E/P distributions in the momentum range $0.4\leq P\leq 4.0$ GeV,
E/P and momentum scatter plots with and without the BCL correction
are shown in
Figure \ref{fig:E/PvsP}.

\section{Detector simulation}

\subsection{Event generation}
For single-photon-exchange process, we used JETSET6.3 as an event generator
\cite{jetset63}.
The parameters used in this program were fitted for general event shapes
\cite{adachi}.
For two-photon processes, we used own event generator of which the details
were described in reference \cite{iwasaki,hayashii,softpi}.

\subsection{Monte-Carlo simulation}
The particles generated by the above-mentioned
event-generation program were traced
through the TOPAZ detector simulator.
This detector simulation was designed to take into account
all conceivable interactions which the particles might experience in
the detector, and to mimic the detector response to them as closely as
possible.
The structures and materials of the detectors are precisely coded in the
program and each particle is propagated by a small step from
the interaction point. At each step, such interactions as decays,
multiple scatterings, and bremsstrahlungs take place according to
the probabilities associated with them.

In the TPC region the averaged energy loss in the gas is calculated
according to a formula determined from the minimum ionizing pions
in real data, knowing the velocity and charge of the particle.
The Landau fluctuation is taken into account by smearing the energy loss
so as to be consistent with the experimental distribution.
The smeared energy loss is translated to a pulse height
on the corresponding wire to be sampled
in order to mimic the digitized outputs.
Because the time profile of the wire signal was
experimentally found to be
independent of the track angle, a single pulse-height
shape function is used to calculate the pulse height in each time bucket.
The pulse height is digitized if it is above the threshold, and is recorded
as wire data.
Pad signals are generated using the pulse heights of the relevant
five wires near to the pad rows.
The avalanches on the five wires induced a charge along the pad row.
We assume that the induced charge is a Gaussian distribution, and is
proportional to the corresponding wire pulse heights. The width of the
Gaussian and the weight factors for the five wires are determined
experimentally \cite{itoh}.

To simulate the electromagnetic shower development in the material,
the EGS code is widely used \cite{egs}.
The cut-off energy in the EGS program, when it is used, must be
set at a value sufficiently lower than a critical energy for an accurate
simulation, which consequently needs more CPU time.
In the BCL, the electromagnetic shower is simulated by using the Bootstrap
method to save CPU time \cite{frozen}.
This method utilizes pre-generated showers to
obtain the shower density in cells of
suitable size in the calorimeter.
The development of a shower with an energy lower than the cut-off is
replaced by them. Thus, once many sets of the frozen showers
are prepared for various energies, the shower simulation can be carried
out as accurately as that of the EGS code, while
at the same time considerably reducing the CPU time.

For a nuclear interaction GEISHA 7 is used \cite{geisha}.
Short-lived particles such as $K_s$, $\pi^0$, and $\Lambda$ are decayed
in the simulator program
while keeping secondary vertex information.

\section{Hadronic-event sample}

\subsection{Trigger}
The charged trigger required at least two charged tracks with
$P_T>$ 0.3-0.7 GeV and an opening angle $>$ 45-70$^o$, depending on the
beam conditions. The neutral energy trigger required that the energy
deposit in the BCL had to be greater than 2-4 GeV, or that in the ECL
had to be greater than 10 GeV. Details concerning the trigger
system can be found in reference \cite{trigger}.

\subsection{Single-photon-exchange process}
Our selection criteria for the single-photon-exchange process are as follows:
(1) the number of good tracks had to be $\geq$ 5, where a good track was
required to have (i) a transverse momentum to the beam axis larger
than 150 MeV, (ii) a distance of closest approach to the beam axis less
than 5cm in the xy-plane, as well as in the z-direction, and (iii) a polar
angle satisfying $0.02\leq |\cos \theta|\leq0.77$ to be well
contained in the detector;
(2) the visible energy ($E_{vis}$) had to be
$\geq$ the beam energy ($E_{beam}$),
where $E_{vis}$ is the sum of the momenta of the
good tracks and the energies of BCL and ECL clusters of energy greater than
100 MeV;
(3) the longitudinal momentum balance had to be
$|\Sigma P_z|/E_{beam}\leq 0.4$, where the $P_z$'s were longitudinal momenta
of the good tracks or longitudinal energies
of the BCL and ECL clusters;
(4) the larger of the hemisphere-invariant masses had to be
greater than 2.5 GeV, where the two hemispheres were divided by a plane
perpendicular to the thrust axis; and
(5) those events with two or more large energy clusters of greater than
0.5$E_{beam}$ were discarded.
In total, 23,783 hadronic events were obtained.
The corresponding luminosity was 197 pb$^{-1}$
and the average center-of-mass energy was 58 GeV.
The details can be found in reference \cite{nakano,miyabaya}.

\subsection{Two-photon process}
The hadronic events produced by two-photon processes were selected based
on the following criteria:
(1) The number of good tracks had to be at least 4, where $|\cos \theta|$
should be less than 0.83;
(2) the position of the origin of the event (i.e., the event vertex),
reconstructed from all tracks, had to be within 1.5cm in the xy-plane and
within $\pm$2.0cm in z-direction;
(3) $E_{vis}$ had to be less than 30GeV, where only the BCL clusters were
used for neutral energy;
(4) the mass of the hadronic system ($W_{vis}$) had to be greater than
3GeV; and
(5) the energy of the most energetic cluster appearing in the BCL,
ECL, and FCL had to be less than 0.4$E_{beam}$
(anti-tag condition).
These selection cuts left 23,779 events at an integrated luminosity of
95.3pb$^{-1}$.
The average center-of-mass energy was 58 GeV.
The details can be found in reference \cite{iwasaki}.

\section{Electron selection for a single-photon exchange}
\subsection{Good track selection}
The electron tracks which satisfied good-track criteria
were selected by the $dE/dx$ information from the TPC
and the E/P ratio measured with the TPC and BCL
in a broad momentum region.

The selection criteria for good tracks were followings:
\begin{enumerate}
\item The closest approach to the beam axis in the xy-plane must be
less than 1.0cm;
\item The closest approach to the interaction point in the z-direction
must be less than 4.0cm;
\item The absolute value of the $\cos\theta$ of the track must be within
between 0.02 and 0.83, where $\theta$ is the polar angle of the tracks
with respect to the initial electron direction;
\item The transverse momentum with respect to the beam axis must be
greater than 0.15 GeV/c;
\item The number of degrees of freedom (N.D.F.) of the track fitting
must be greater than 3, with a maximum of 15;
\item The number of wires for the $dE/dx$ calculation (65-\% truncated
mean) must be greater than 30, with a maximum of 114; and
\item The momentum must be greater than 0.8 GeV/c.
\end{enumerate}

Conditions (1),(2), and (5) require good measurements of the
momentum of the tracks. (1) also reduces the background.
Condition (3) requires the tracks to be within the TPC active volume.
(4) requires that the tracks are not curled up in the TPC volume.
(6) is the condition for the sufficient resolution of $dE/dx$.
(7) is added to reduce the background.

\begin{figure}
\vskip -2cm
\epsfysize10cm
\hskip1in\epsfbox{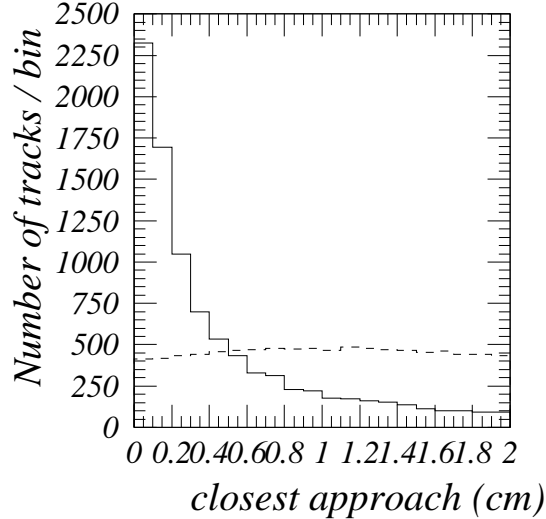}
\vskip -1cm
\caption{
Distributions of the closest approach in the Monte-Carlo
simulation; the solid histogram is for prompt electrons and the dashed
one is for pair conversion and Dalitz decays.
}
\label{vertex}
\end{figure}

\begin{figure}
\vskip -2cm
\epsfysize10cm
\hskip1in\epsfbox{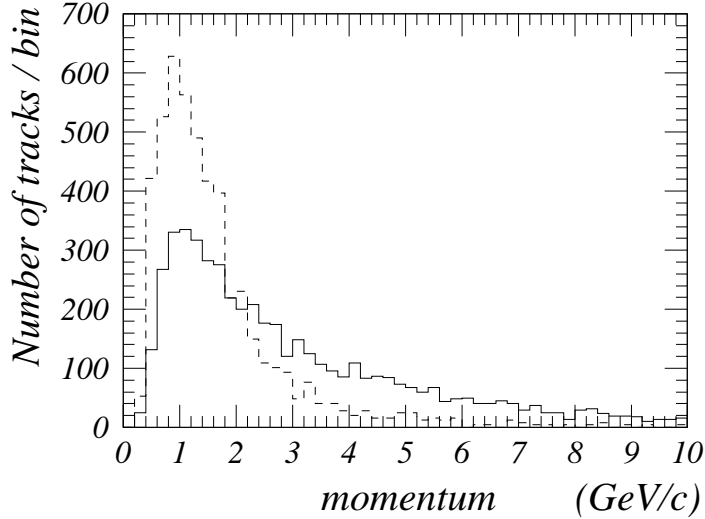}
\vskip -1cm
\caption{
Distributions of the momentum in the Monte-Carlo
simulation; the solid histogram is for prompt electrons and the dashed
one is for the background.
}
\label{momdis}
\end{figure}

Figure \ref{vertex} shows the difference of the closest approach to the
beam axis in the xy-plane for prompt electrons and the
background tracks in a Monte-Carlo simulation.
Figure \ref{momdis} shows the difference in the momentum distribution
for prompt electrons and the background tracks in a
Monte-Carlo simulation.

After the tracks were
subsequently extrapolated to the radius of each BCL cluster center,
we calculated the distance to the clusters.
The cluster was matched with the track having the closest distance.
E/P was calculated using a matched track and cluster.

\subsection{Electron selection}
The electron selection was carried out based on two criteria.

First, we calculated the $\chi^2_i$s for each hypothetical particle using a
$dE/dx$ information and a momentum measurement from the TPC, where $i$
indicates an electron, muon,pion, kaon, or proton. The definition of
$\chi^2_i$ is
$$
\chi^2_i = \left({\frac{\frac{dE}{dx}_{meas.}-\frac{dE}{dx}(\beta\gamma)}
{\Delta\frac{dE}{dx}_{meas.}}}\right)^2
+\left({\frac{\beta\gamma^i_{meas.}-\beta\gamma}
{\Delta (\beta\gamma)_{meas.}}}\right)^2,
$$
with
$$
\beta\gamma^i_{meas.} = \frac{P_{meas.}}{M_i}.
$$
Here, $\Delta\frac{dE}{dx}_{meas.}$ and $\Delta (\beta\gamma)_{meas.}$ are
the estimated errors of $\frac{dE}{dx}_{meas.}$ and
$\beta\gamma^i_{meas.}$, respectively. They include inefficiency
measurements in the TPC.
$\Delta\frac{dE}{dx}_{meas.}$ also includes the effect of the nonlinear shift
described before.
The most probably $\chi^2_i$ is given by minimizing it while
modifying $\beta\gamma$.
The expected $dE/dx$ is given by a polynomial
function of $\beta\gamma$.
The details are described in Section \ref{bethe}.

\begin{figure}
\vskip -2cm
\epsfysize10cm
\hskip1in\epsfbox{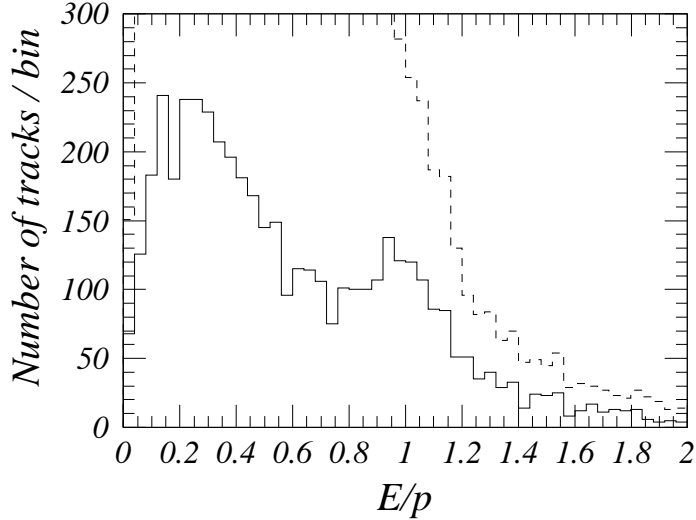}
\vskip -1cm
\caption{
E/P distribution of the experimental data; the dashed
histogram is for the before $dE/dx$ selection and the solid one is for
the after $dE/dx$ selection.
}
\label{epsel}
\end{figure}

The electron-candidate tracks were selected by requiring that $\chi^2_e$
must be less than 3.0, where the number of degrees of freedom (N.D.F.)
is 1.
Figure \ref{epsel} shows the E/P distributions
both before and after this selection.
The peak of the electrons is enhanced.

The rejected tracks in this selection ($dE/dx$-hadron) are used for
hadron background estimations.

\begin{figure}
\vskip -2cm
\epsfysize10cm
\epsfbox{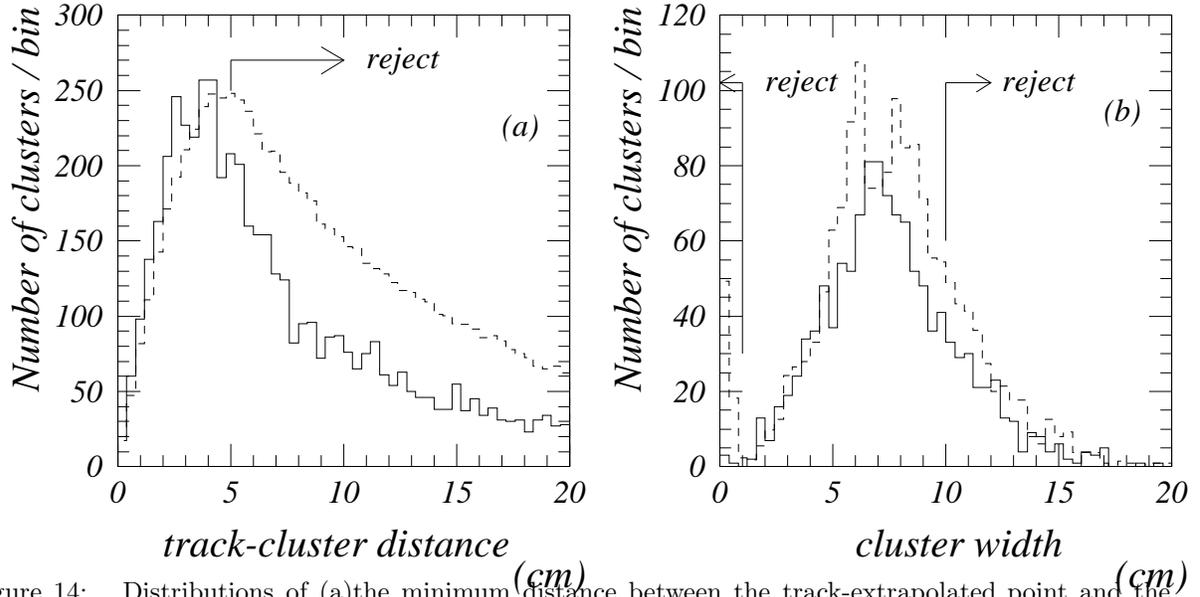}
\vskip -1cm
\caption{
Distributions of (a)the minimum distance between the
track-extrapolated point and the cluster, and (b)the shower width
with respect to the track; the solid histograms and the dashed
histograms are the total tracks and the electron-enhanced samples of the
hadronic events, respectively.
}
\label{width}
\end{figure}

Second, we required the electron-candidate tracks to have a
distance of less than 5.0cm with respect to the calorimeter cluster positions,
and calculated the cluster width, which
is the energy-weighted r.m.s. of counter positions with respect to the
extrapolated track position. We also carried out selection for the
cluster width, i.e., we selected the clusters having widths of between
1.0cm and 10.0cm. These distributions for electron-candidate tracks and
hadron tracks are shown in Figure \ref{width}.

\begin{figure}
\vskip -2cm
\epsfysize10cm
\hskip1in\epsfbox{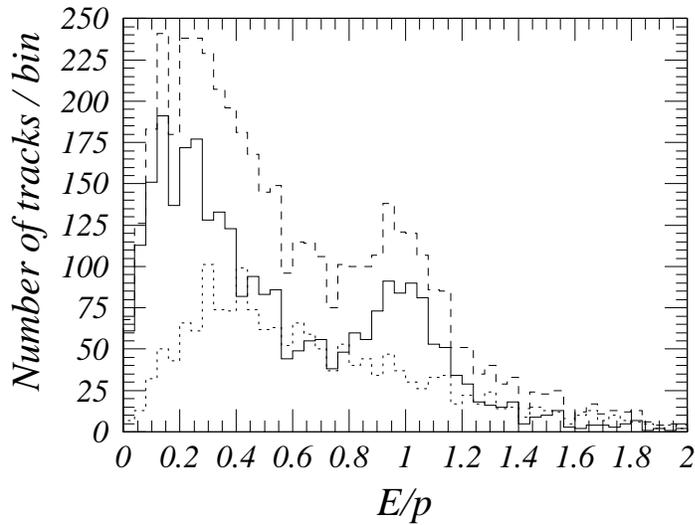}
\vskip -1cm
\caption{
E/P distribution of the experimental data; the solid
and the dashed histograms are after and before the shower selection,
respectively. The dotted one is for the rejected tracks by this
selection.
}
\label{shcut}
\end{figure}

The E/P distributions of before and after these selections are shown in
Figure \ref{shcut}.

\subsection{Background estimation}

\subsubsection{$\gamma$-conversion rejection}
Since the largest background source
was electrons from
$\gamma$-conversions at the material in front  of the TPC,
we rejected the dominant part of such  electrons by the following
methods.

We reconstructed secondary vertices ($V^{0}$'s)
with an electron-candidate track and one of the
unlike sign charged tracks, and
then calculated the invariant mass of each $V^{0}$, assuming that its
daughter particles are electrons.
For the $V^{0}$-reconstruction,  two
kinds of vertices, i.e.,  non-crossing and crossing cases in the xy-plane
(perpendicular to the beam axis), were searched.
In the former case,
the combinations with a distance in the xy-plane
of less than 4.0cm and a z-distance of less than 2.0cm were
selected; we required that
the deflection angle between the momentum vector of two tracks and the
position vector of the closest point be less than 5.0 degrees.
In the latter case,
we selected one vertex which had
a smaller deflection angle, and required that
the z-distance be less than 2.0cm and the
deflection angle be less than 5.0 degrees.

We then rejected the tracks in the pair if its invariant mass was
 $\leq$ 50 MeV in the former or $\leq$ 200 MeV
in the latter cases, respectively.

\begin{figure}
\epsfysize15cm
\epsfbox{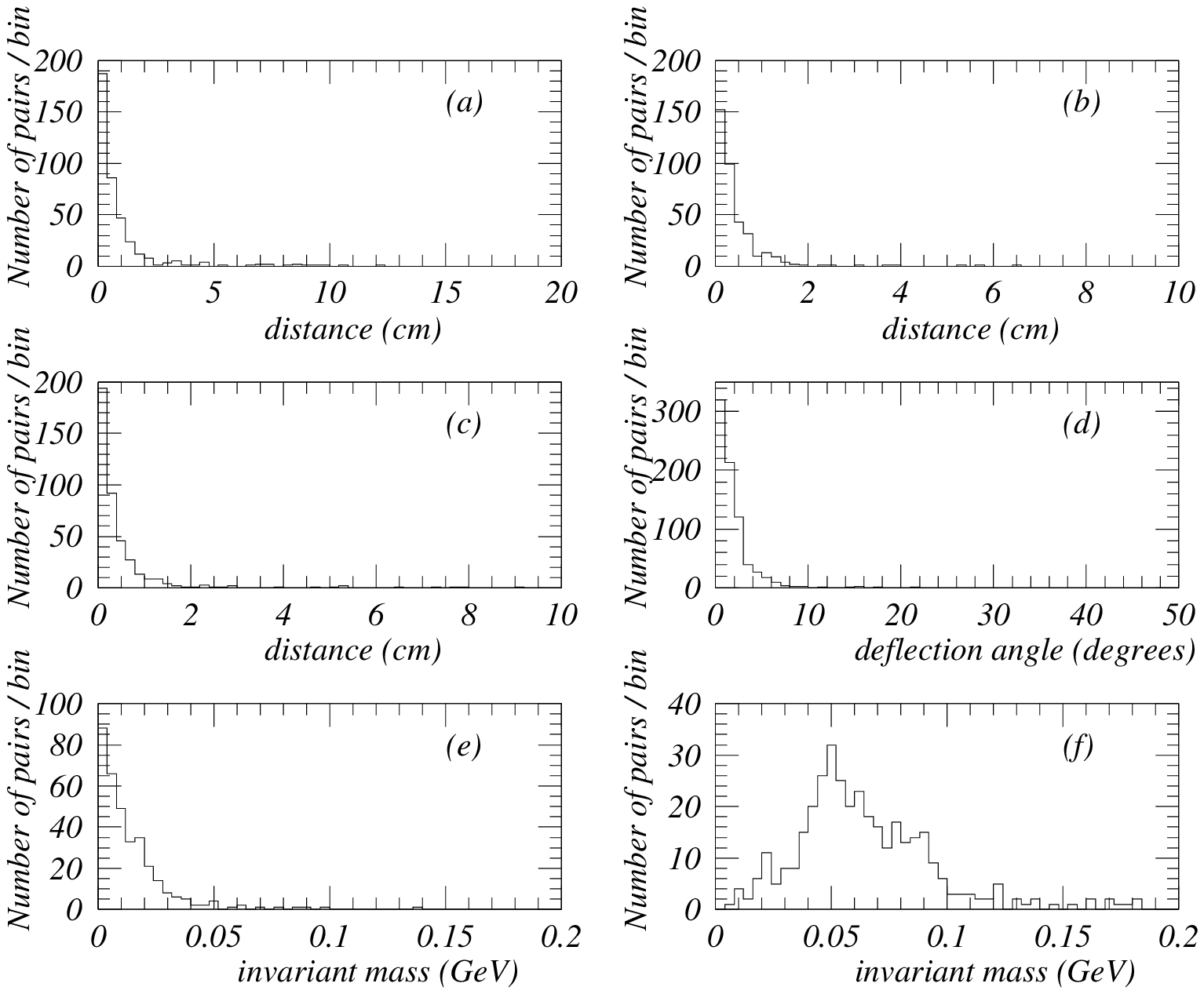}
\caption{
Distributions of the secondary vertices in the
Monte-Carlo simulation; (a)the distance in the xy-plane of the not
intersected case, (b)the distance in the z-direction of the not
intersected case, (c)the distance in the z-direction of the
intersected case, (d)the deflection angle of both cases, (e)the
invariant mass of the not-intersected case, and (f)the invariant mass
of the intersected case.
}
\label{svutl}
\end{figure}

These distributions are shown in Figure \ref{svutl}.

\subsubsection{Hadron background estimation}
\begin{figure}
\vskip -2cm
\epsfysize10cm
\hskip1in\epsfbox{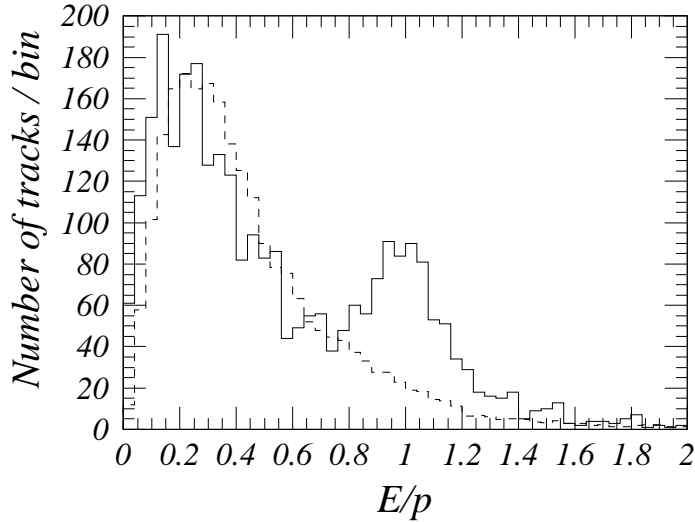}
\vskip -1cm
\caption{
Example of the hadron background estimation; the solid
and dashed histograms are the selected electron candidates and the
$dE/dx$-hadrons which is described in the text,
respectively.
}
\label{eop}
\end{figure}

The hadron background on the E/P plot
 was estimated using the $dE/dx$-hadrons. The distributions were
normalized using the entries between 0.0 and 0.64. The normalized E/P
distributions of the $dE/dx$-hadrons were subtracted from the E/P
distributions of the selected electron candidates. The numbers of
electrons were obtained by counting the entries between 0.74 and 2.0
in the subtracted E/P distributions. An example of the
hadron-background estimation is shown in Figure \ref{eop}.

\subsubsection{Remaining $\gamma$-conversion estimation}
\label{section:estimation}
There remained electrons resulting from $\gamma$-conversions
which escaped from $V^{0}$  reconstruction.
They were presumably energy-unbalanced pairs,
for each of which the lower $P_T$ track was not
reconstructed by  the TPC.
We estimated the failure rate of the $V^{0}$
reconstruction by a Monte-Carlo simulation.
The failure rate,
  $ \eta = N^{M.C.}_{V^{0}failure}/N^{M.C.}_{V^{0}reconstructed}$,
was estimated for each $P_T$-bin.
In the calculation, we included contributions from
 Dalitz decay($\pi^{0} \rightarrow e^+ e^- \gamma$) as well
as the conversion electron pairs.
Since we did not apply any vertex position cut in the conversion-pair
reconstruction, this factor  included the effect on
the Dalitz pairs.
In order to estimate the number of remaining background tracks from
$\gamma$-conversions, we multiplied
the number of reconstructed conversion pairs in the experiment
by $\eta$ in each $P_T$ bin.

\subsection{Pion-rejection factor, electron efficiency}
The electron selection in the single-photon-exchange process is summarized
here. The pion-rejection factors were derived using single-track
multi-hadron events generated by Monte-Carlo simulations.
In order to derive the efficiency and the factor, we required E/P
to be within 0.72 to 2.0 for the selected tracks.
All of the tracks which passed the above criteria were counted.
The definition of the selection efficiency is the ratio of the number
of electrons counted (as above) to the number of
charged tracks
before all of the selection criteria.
The pion-rejection factor is defined as the ratio of the number of
pions before selection to the number of the remaining pions with
an E/P of between 0.72 and 2.00.

\begin{figure}
\vskip -2cm
\epsfysize10cm
\hskip1in\epsfbox{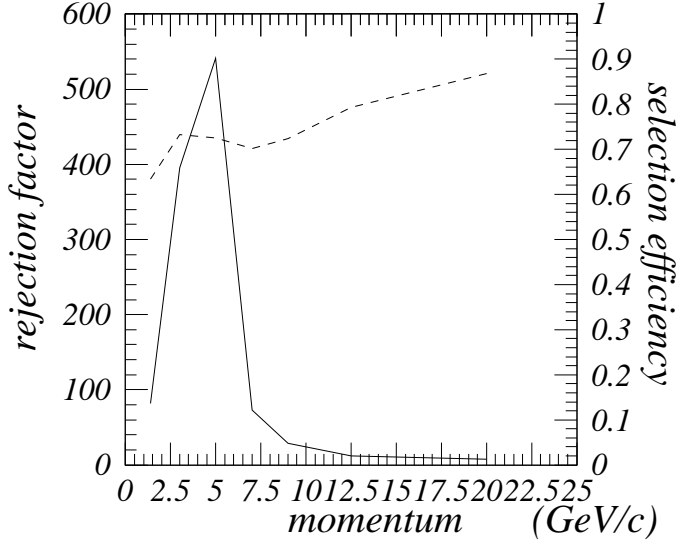}
\vskip -1cm
\caption{
Pion-rejection factor and electron efficiency for
single-track events;
the solid curve is the pion-rejection factor (left scale)
and the dashed one is
the electron efficiency (right scale).
}
\label{single}
\end{figure}

This value for single-track events is shown in Figure \ref{single},
together with the electron efficiency.
Since the $dE/dx$ curve of the pions is close to that of the electrons at
the higher momentum region, the pions satisfy the $dE/dx$ selection
criterion. Thus, the pion rejection factor is worse in the higher
momentum region.
The average values were 80 for pion rejection
and 68.4\% for electron efficiency.
In order extract these two averages, we assumed the momentum spectrum
in the multi-hadron events.

In practice, further rejection was carried out.
The pions remaining after the E/P selection were estimated using the
tracks rejected by the $dE/dx$ selection ($dE/dx$-hadrons). The E/P
distributions of the $dE/dx$-hadrons were normalized using the entries in
the side-band ($0.0<$E/P$<0.64$); then the distribution from
the E/P distribution of the selected tracks
was subtracted.
This procedure improves the pion rejection factor by more than factor 3.

\begin{figure}
\vskip -2cm
\epsfysize10cm
\hskip1in\epsfbox{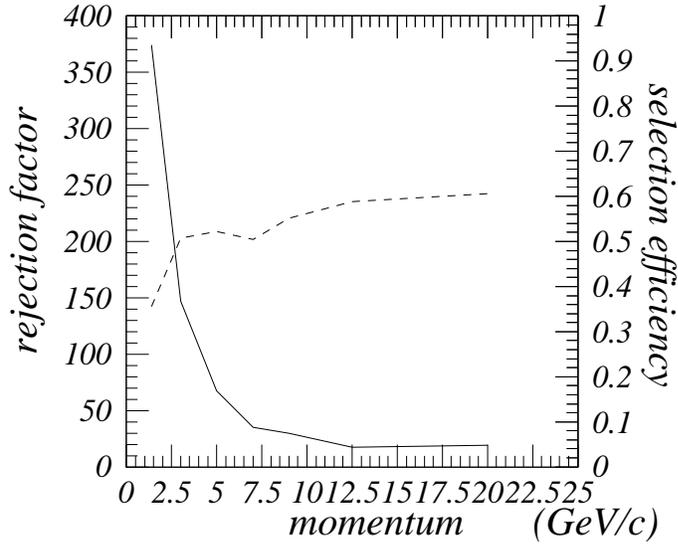}
\vskip -1cm
\caption{
Pion-rejection factor and electron efficiency for low-$P_T$
tracks; the solid curve is the pion-rejection factor (left scale)
and the dashed one is
the electron efficiency (right scale).
}
\label{lowpt}
\end{figure}

\begin{figure}
\vskip -2cm
\epsfysize10cm
\hskip1in\epsfbox{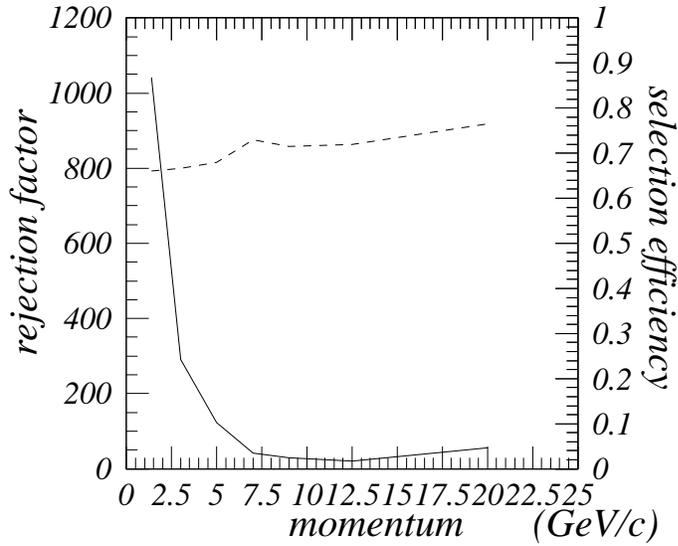}
\vskip -1cm
\caption{
Pion-rejection factor and electron efficiency for high-$P_T$
tracks; the solid curve is the pion-rejection factor (left scale)
and the dashed one is
the electron efficiency (right scale).
}
\label{highpt}
\end{figure}

For multi-hadron events,
we show these values for two cases, i.e., low-$P_T$ ($P_T<0.8$ GeV) and
high-$P_T$ ($P_T>0.8$ GeV). They are shown as a function of the electron
momentum in Figures \ref{lowpt} and \ref{highpt}.
The same E/P selection was carried out
as that described above to derive these
Figures.
For the higher momentum tracks, the pion-rejection factors are rather
worse because of the $dE/dx$ effect (as described before).
To summarize, the pion-rejection factors of high-$P_T$ tracks and
low-$P_T$ tracks were estimated to be 164 and 137 on the average,
where the electron efficiencies were 68.4 and 42.7\%, respectively.
Practically, we also used additional pion rejection by fitting
(as mentioned above). We can thus
we can expect better rejection of the pions.

\section{Electron identification in two-photon processes}

\subsection {Electron selection}
In two-photon processes electron tracks were identified
by combining the information from the E/P ratio, $dE/dx$,
and the TOF as follows:
\begin{enumerate}
\item Charged tracks in the TPC were extrapolated to the BCL.
We then selected for each TPC track the BCL cluster
which was the closest. The E/P of each of the so-selected TPC-BCL
combinations had to satisfy $0.75 \leq$ E/P $\leq 1.25$.
\item $dE/dx$ was calculated for the electron-track candidate
to be used in the subsequent electron counting.
The resultant $dE/dx$ was mostly contained
in the range $5.5 \leq dE/dx \leq 7.5$ keV/cm.
\item
We used information from the TOF and required the track to have
a confidence level(CL) of 0.01 or better for the electron hypothesis.
\end{enumerate}
 The energy-loss
($dE/dx$) information from the TPC enables us to separate
electrons from hadrons in the momentum region($P_{T} < 3$GeV).
The E/P ratio of the energy(E) measured by the BCL
and the momentum(P) measured by the TPC can separate
electrons clearly above 0.4 GeV.
The TOF was useful to resolve electrons from kaons and protons in the
 overlapping region of $dE/dx$.

\begin{figure}
\epsfysize18cm
\hskip-1cm\epsfbox{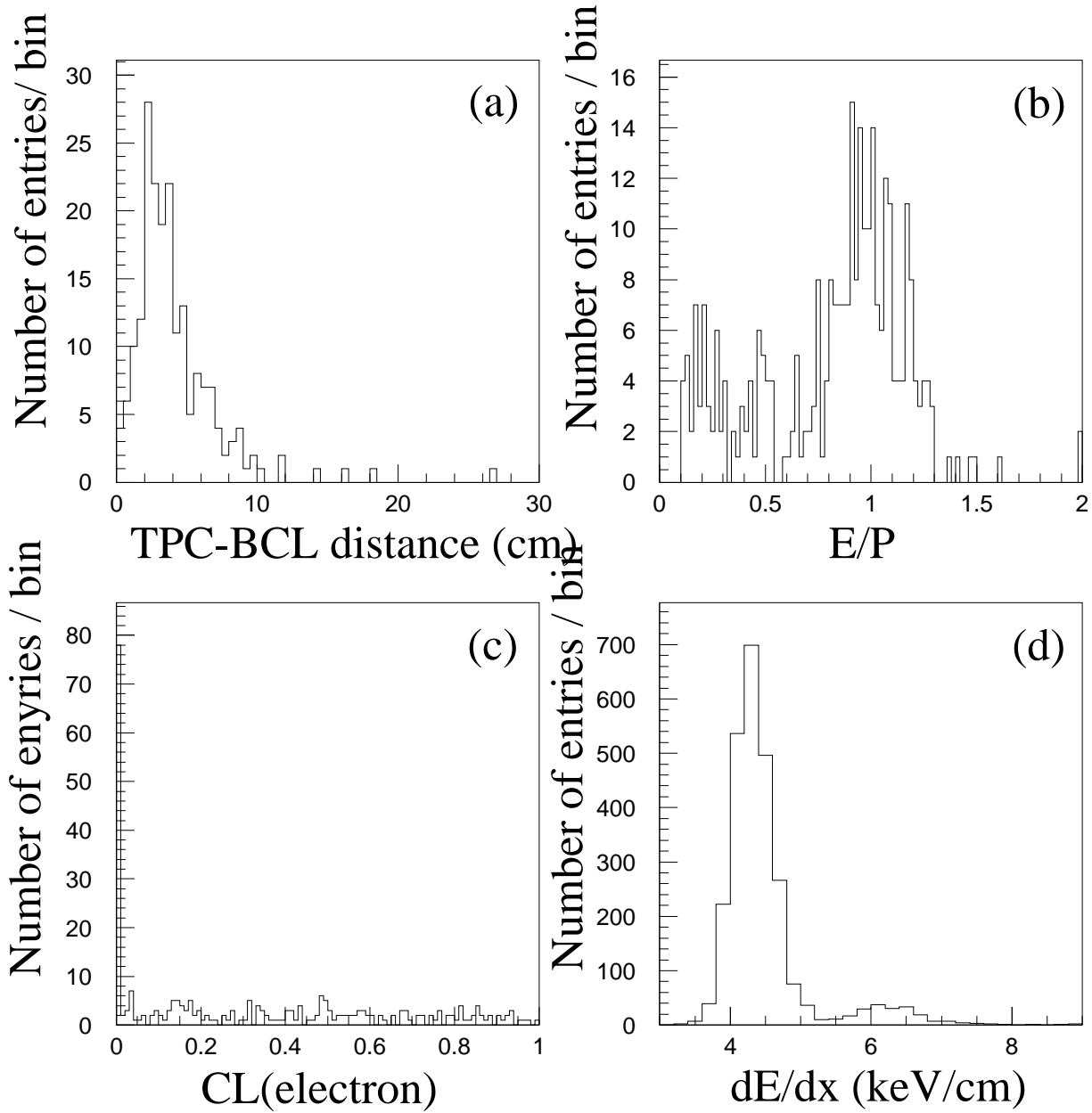}
\caption{
Various distributions for electron candidates:
(a) the distance between a TPC track and
its closest BCL cluster,
(b) E/P, (c) the confidence level (CL) for the TOF, and (d) $dE/dx$.
Each of these distributions was obtained
with all of the cuts except for that on the plotted quantity.
}
\label{fig:selection}
\end{figure}

The performance of electron identification in the momentum range 0.4
$\leq P_{T} \leq$ 3.0 GeV and the polar-angle range $|\cos\theta|\leq$
0.77 is demonstrated
in  Figures \ref{fig:selection},
where various distributions
(the closest TPC-BCL distance(Figure \ref{fig:selection}-(a)),
the E/P ratio(Figure \ref{fig:selection}-(b)),
the CL for the electron hypothesis in the TOF(Figure \ref{fig:selection}-(c)),
and the $dE/dx$(Figure \ref{fig:selection}-(d))) are shown.
In these figures, electrons coming from $\gamma$-conversions were
already rejected. The methods for $\gamma$-conversion rejection are
described in the next section.
Notice that these figures were obtained
with the electron candidates selected by all of
the cuts except for that on each plotted quantity.
Notice in particular that Figures \ref{fig:selection}-(a) to (c) were
made with an $dE/dx$ cut ($5.5 \leq dE/dx \leq 7.5$ keV/cm) which was
used only for the purpose of displaying.
 As can be seen from the $dE/dx$ distribution in Figure
\ref{fig:selection}-(d),
two peaks corresponding to electrons and pions are clearly separated.

\begin{figure}
\epsfysize18cm
\epsfbox{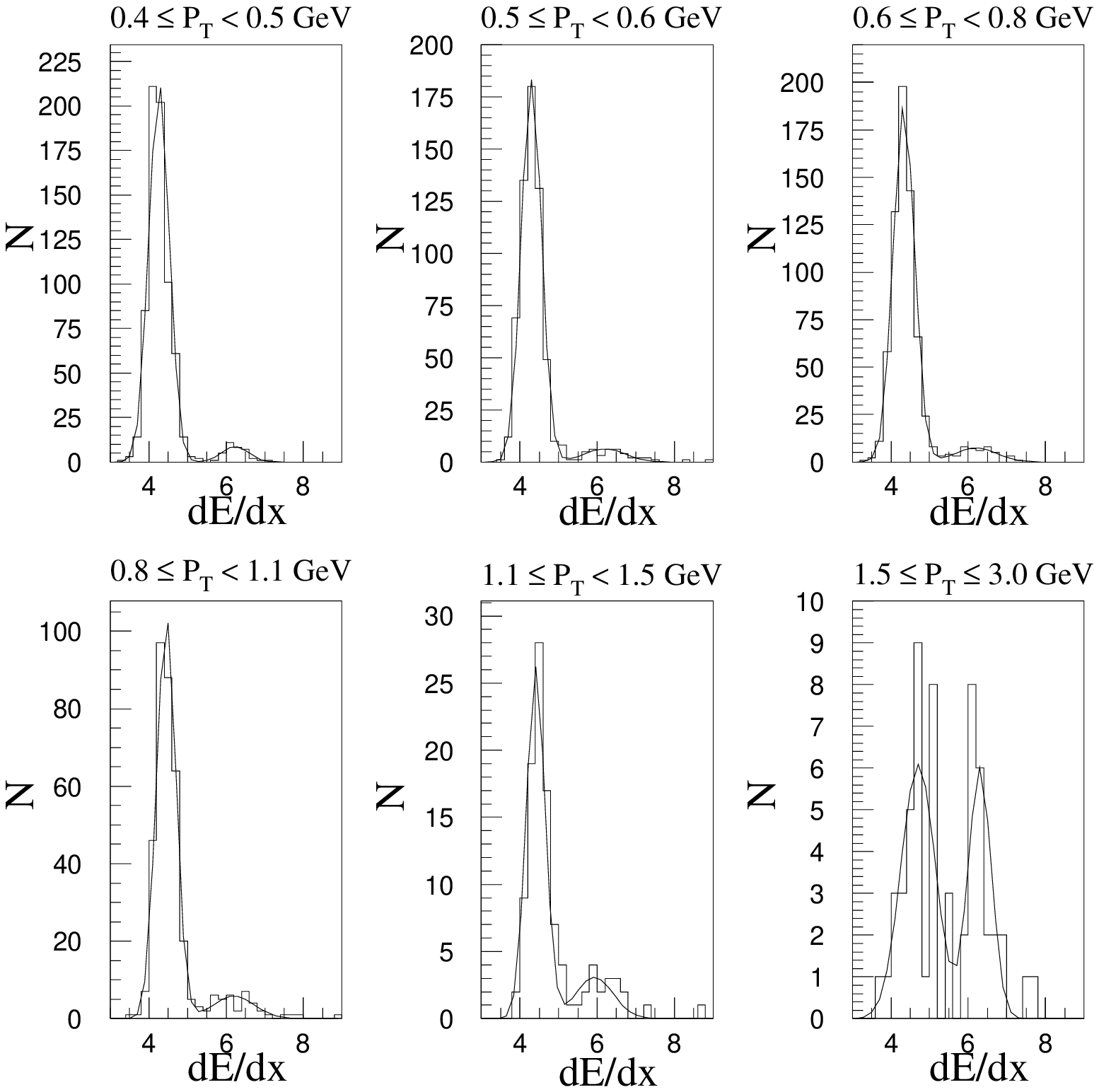}
\caption{
$dE/dx$ distributions for each $P_T$-bin indicated in the figure.
The smooth lines are
the best-fit curves described in the text.
}
\label{fig:dedx}
\end{figure}

We counted the numbers of electrons in each $P_{T}$ bin by
fitting the $dE/dx$ distributions with double Gaussians bin by bin.
The $P_T$ binning was selected so as to approximately equalize
the number of entries in each bin.
The $dE/dx$ distributions are shown in Figures  \ref{fig:dedx}.

\subsection{Reject $\gamma$-conversions}
The electrons from the $\gamma$-conversions
were rejected by secondary vertex ($V^{0}$) reconstruction.
The $V^{0}$'s were reconstructed from all combinations of two tracks,
and the invariant mass of each $V^{0}$ were calculated.
When two tracks did not intersect at the closest point in the
xy-plane,  the distances of the two tracks
at the minimum-distance position were required
to be less than 7 cm in the xy-plane and 3 cm in the z-direction.
When two tracks intersected, we chose from the two crossing points
that with the shorter z-difference, and required it to be
less than 1.5 cm.
We then rejected the tracks in the pair if its invariant mass was
 $\leq$ 80 MeV in the former or $\leq$ 150 MeV
in the latter cases.

We also required the closest approach of each electron-track candidate
to the event vertex in the xy-plane to be $<$ 0.5-1.5 cm, depending on
$P_{T}$.

\subsection{Remaining $\gamma$-conversion estimation}
The remaining $\gamma$ conversions and Dalitz decays were
estimated using the failure rate of the $V^{0}$
reconstruction which was obtained by a Monte-Carlo simulation.
The failure rate of the $V^{0}$ reconstructions
(described in section \ref{section:estimation})
was estimated for each $P_T$-bin, and was typically $\sim$0.4.
The number of remaining background tracks
was estimated by this ratio multiplied by
number of the reconstructed conversion pairs in the experiment
in each $P_T$ bin.
The $\gamma$-conversion and Dalitz decay background mainly occupy the
low-$P_T$ region.

\subsection{Pion rejection factor and electron efficiency}
\begin{figure}
\vskip -2cm
\epsfysize10cm
\hskip1in\epsfbox{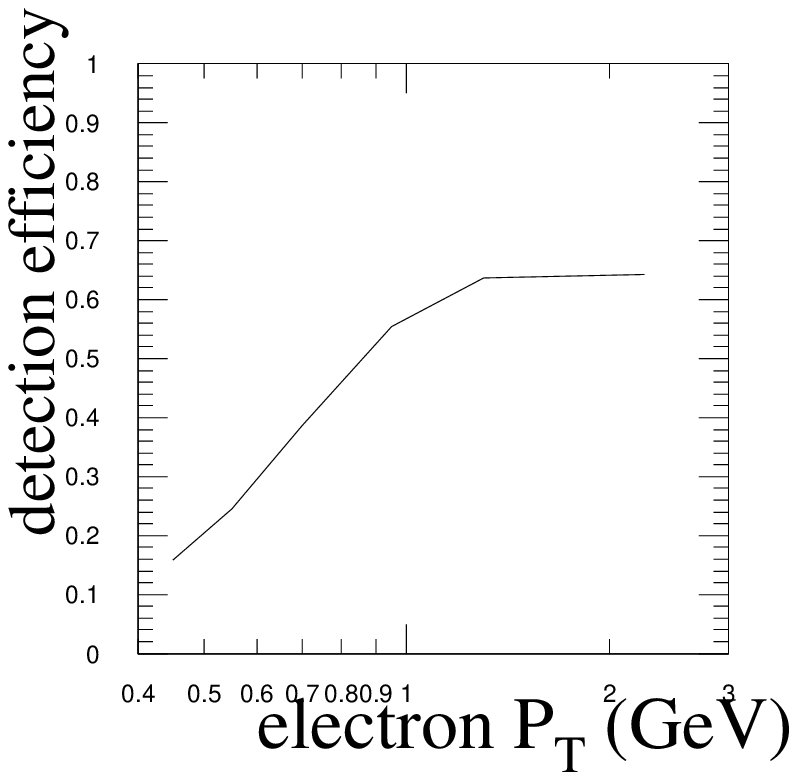}
\vskip -1cm
\caption{
Detection efficiency of electron tracks for
single-track events.
}
\label{single2}
\end{figure}

\begin{figure}
\vskip -2cm
\epsfysize10cm
\hskip1in\epsfbox{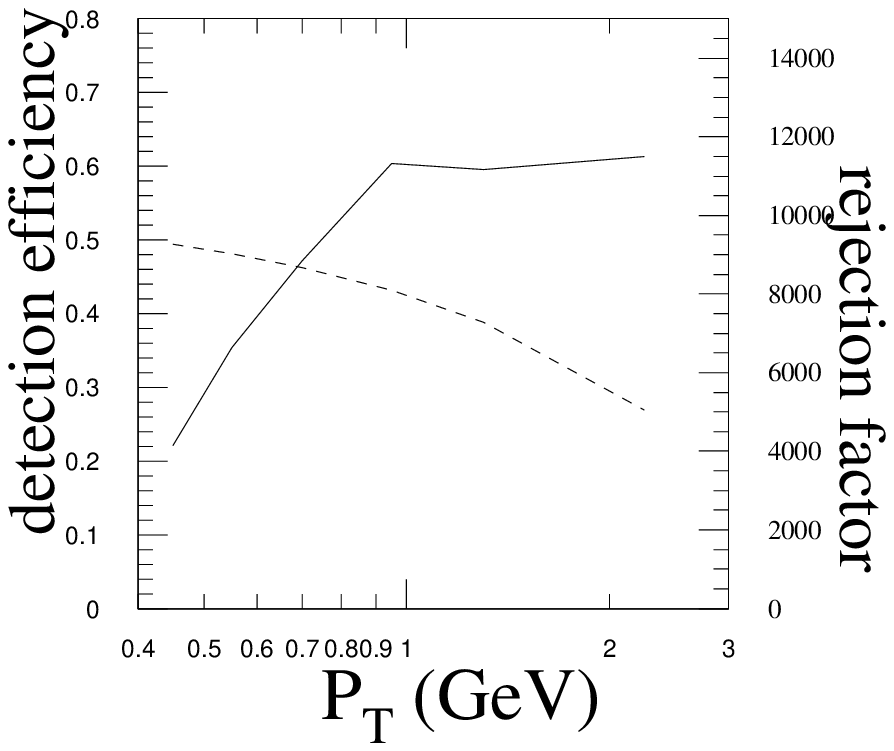}
\vskip -1cm
\caption{
Pion-rejection factor and electron efficiency for
two-photon hadronic events;
the dashed curve is the pion-rejection factor (right scale)
and the solid one is
the electron efficiency (left scale).
}
\label{twophoton}
\end{figure}

The electron selection in the two-photon processes is summarized
here.
The electron efficiencies for single-track events and for
multi-hadron events which were generated by the Monte-Carlo simulation
are shown in Figures \ref{single2} and \ref{twophoton},
respectively, as a function of the transverse momentum of the track.
The average values of the efficiencies were 34.2\% for single-track events
and 36.0\% for multi-hadron events.
The former value was obtained assuming the $P_T$ distribution in the
multi-hadron events.

The pion-rejection factors were also derived with single-track
and multi-hadron events.
For single-track events we obtained a pion-rejection
factor of 15000 for the high momentum region of $1.5 \le P_T \le 3.0$ GeV
where the factor was the smallest.
The pion-rejection factor for multi-hadron events is indicated
in Figure \ref{twophoton} by the dashed line. The average value for
multi-hadron events was 8600.
In the real analysis we counted the number of electrons by fitting
the $dE/dx$ distributions.
Therefore, in fact, the pion-rejection factor was far better than
these values.

\section{Summary}
We have developed a method to identify electrons inside hadronic jets
using the time-projection chamber and the lead-glass calorimeter
in the TOPAZ detector system at TRISTAN.
The energy-loss measurement with the time-projection chamber and the
energy measurement with the lead-glass calorimeter provided good
electron identification over a wide momentum range.
In particular, low-$P_T$ ($P_T>0.4$ GeV) electron identification was
proven to be possible. This is powerful for studying the open-charm
production in two-photon processes.
The pion-rejection factors and the electron efficiencies
inside hadronic jets were obtained to be
163 and 68.4\% for
high-$P_T$ electrons and 137 and 42.7\% for low-$P_T$ electrons in
single-photon-exchange process, and
8600 and 36.0\% for two-photon processes.

\section*{Acknowledgement}
We thank the members of the TOPAZ collaboration for their support
and valuable discussions. Special thanks go to Drs. R. Itoh, K. Nagai,
and T. Tauchi for the analysis and calibration for the time-projection
chamber and the lead-glass calorimeter.
We appreciate all of the engineers and technicians at KEK as well as
the collaborating institutions.
This work was partially supported by a Grant-in-Aid for Scientific
Research from the Japan Ministry for Education, Science and Culture.
One of us (E.N.) was supported by Japan Society for the Promotion
of Science.

\end{document}